\documentclass[prd,showpacs,showkeys,preprintnumbers,aps]{revtex4}
\usepackage[dvips]{graphicx}
\usepackage{amsmath,amssymb}
\makeatletter
 \def\setcaption#1{\def\@captype{#1}}
\makeatother

\begin{document}

\preprint{TH-1026}
\title{Color Ferromagnetic Quark Matter in Neutron Stars}

\author{Aiichi Iwazaki$^{\rm a}$\footnote{$\!\!$Electronic address: \tt 
iwazaki@yukawa.kyoto-u.ac.jp},
Osamu Morimatsu$^{\rm b}$\footnote{$\!\!$Electronic address: \tt 
osamu.morimatsu@kek.jp}, 
Tetsuo Nishikawa$^{\rm c}$\footnote{$\!\!$Electronic address: \tt
nishi@th.phys.titech.ac.jp} and
Munehisa Ohtani$^{\rm d}$\footnote{$\!\!$Electronic address: \tt
ohtani@rarfaxp.riken.jp}}

\address{$^{\rm a}$Department of Physics, Nishogakusha University, 
Kashiwa Ohi Chiba 277-8585,\ Japan. \\
$^{\rm b}$Institute of Particle and  Nuclear Studies, High Energy
  Accelerator Research Organization, 1-1, Ooho, Tsukuba, Ibaraki,
  305-0801, Japan \\
$^{\rm c}$Department of Physics, Tokyo Institute of Technology, Meguro,
  Tokyo, 152-8551, Japan\\
$^{\rm d}$Radiation Laboratory, RIKEN,
2-1 Hirosawa, Wako, Saitama 351-0198, Japan}

\date{July 11, 2005} 
\begin{abstract}
Color superconducting state has been known as a possible phase
of quark matter with sufficiently large baryon number
density so as for perturbative analysis to be valid. 
We point out that 
a color ferromagnetic state is another possible
phase of such a sufficiently dense quark matter.
Furthermore,
we show that the color ferromagnetic phase 
is energetically more favored than the color superconducting phase
in the quark matter with smaller baryon number density.  
We expect that increasing baryon density in neutron stars
transforms nuclear matter into the quark matter
of the color ferromagnetic phase, not color superconducting phase.
We find that a critical mass of the 
neutron star with such an internal structure is
about $1.6M_{\odot}$. 
\end{abstract}
\pacs{12.38.-t, 12.38.Mh, 25.75.Nq, 26.60.+c, 97.60.Jd}
\keywords{Quark Matter, Color Superconductivity, Quantum Hall States,
Neutron Stars}
\maketitle
%
%
Neutron stars have quite dense nuclear matter\cite{neutron} 
at their inner core.
The density is expected to reach at least several times the normal nuclear 
density ($\simeq 2.8\times 10^{14}\mbox{g/cm}^3$). 
Observationally\cite{ob}, the average density of
the neutron stars is about twice the normal nuclear density, 
when their mass is $1.4 M_{\odot}$ and their radius is $10$km.
(Even if we take the radius as $13$km, the average density
exceeds the normal nuclear density.)
Although their radii have not yet been determined definitely, 
this naive estimation implies that the hadronic matter at the center of 
the neutron stars is much more dense than the normal nuclear density.

It is natural to expect that the neutron stars 
involve quark matter at their center. 
As the density of the hadronic matter increases, nucleons begin to 
overlap each other at about three times the normal nuclear density.
This suggests that a transition between hadrons and quarks 
occurs at a few times the normal nuclear density and
hence the quark matter may be realized at a core of the neutron stars.
Thus, it is quite important to analyze the quark matter in 
understanding properties of the neutron stars.

So far, several phases of the quark matter have been studied: normal 
quark gas state, color superconducting state\cite{color} and color 
ferromagnetic state\cite{cf,cf3} which we have recently pointed out. 
In the normal quark gas state, quarks interact weakly with
each other and 
the SU(3) gauge symmetry remains exact.
This state is naively expected to be realized in
quark matter with sufficiently large baryon number density
so as for the gauge coupling constant to be much small.
But this expectation does not hold in the quark matter with low
temperature. Such a quark matter has been shown to
form di-quark condensates and form  
the color superconducting states; color flavor locked (CFL) state involves three flavors in the
condensates,
while two-flavor superconducting (2SC) state does only two light flavors in them.
Owing to these condensates, the SU(3) gauge symmetry is
broken fully or partially in the color superconducting states. 
On the other hand,
the color ferromagnetic state (CF)
is such a state that a color magnetic field
arises spontaneously due to gluon's dynamics
and quarks occupy Landau levels under the color magnetic field.
Because of the magnetic field, the gauge symmetry is partially broken also
in this state.

The color ferromagnetic state was firstly discussed by Savvidy\cite{savvidy} 
as a more stable vacuum than the perturbative one, namely, as a possible
candidate for
quark matter with
vanishing baryon number density.
This argument is based on the fact that
a nonzero color magnetic field minimizes
the one-loop effective potential of the magnetic field.
Since the one-loop approximation
is valid in a system with small gauge coupling constant,
this CF state may be realized
in the quark matter 
with sufficiently large baryon number density, not 
in the quark matter with vanishing baryon number density.
Since the energy density of quarks
under the color magnetic field is lower than
that in the absence of the field, the nontrivial minimum remains
in the effective potential including the effects of the quarks.
Therefore, the CF is a possible phase of
the dense quark matter as well as the color superconducting states.

There is, however, a naive argument that
the color magnetic field does not arise 
because magnetic mass of gluons
could be produced in the dense quark matter;
owing to the mass, the magnetic field is expelled or
screened. We would like to mention in this point that
although the electric mass, i.e.~Debye mass, of the gluons can be
generated in the dense quark matter, 
the magnetic mass can be hardly produced just as
the magnetic mass of photons in electron gas. Only the exception is  
the color superconducting quark matter\cite{huang}, where
the magnetic field is suppressed.

This nontrivial state, however, was shown\cite{nielsen} to be unstable
in the vacuum soon after the work by Savvidy.
There exist unstable modes of gluons which possess imaginary mass.
We have shown\cite{cf,cf3} that the instability can be cured 
(see also \cite{bor}) in the dense quark matter by
the formation of fractional quantum Hall states 
of the unstable gluons. 
Therefore, there are two possible phases of the dense
quark matter at low temperature; the color superconducting phase without
the color magnetic field and the stable ferromagnetic phase with the
color magnetic field.
In the color superconducting phase, the energy density
of quarks is smaller than that of the normal quark gas 
due to the di-quark condensates. Similarly, in the color ferromagnetic phase,
their energy density is smaller than that of
the normal quark gas due to the presence of the magnetic field.
Obviously, these two phases are incompatible with each other.

In this letter we determine which phase is realized in the quark matter
with sufficiently large baryon density 
so as for the perturbative approximation or loop approximation to 
be valid.  
For the purpose, we compare the free energy of the phases
at zero temperature 
in terms of baryon chemical potential and find
the state with the lowest free energy. 
The free energy, $\Omega$, of quark matter has contributions from quarks 
and gluons,
\begin{equation}
\Omega=\Omega_q+\Omega_{gl}.
\end{equation}    
where $\Omega_q$ represents the free energy of quarks, while $\Omega_{gl}$ does
vacuum energy in which main contributions come from gluon's loops (and gluon's condensation
in the case of CF phase).

We first estimate
the vacuum energies, $\Omega_{gl}$ in each phase of
sufficiently dense quark matter.
In the one-loop approximation,
the gluon vacuum energy \cite{cf,cf3}, 
${\rm Re}V(gB)=\frac{9}{32\pi^2}(gB)^2(\log(gB/\Lambda^2)-1/2)$
vanishes at $gB=0$ in the
color superconducting phases, while it takes 
${\rm Re}V(gB=\Lambda^2)=\frac{9}{64\pi^2}(gB)^2$ in CF phase;
we have included loop effects of quarks with three flavors in $V(gB)$.
($g$ is a gauge coupling constant.)
We should also take into account of the condensation energy of
gluons, which stabilize the color magnetic field by forming
a quantum Hall state in the CF phase. The energy is
given by $-2B^2=-2(gB)^2/g^2$ \cite{cf3}.
Thus, the free energies, $\Omega_{gl}$, are given by

\begin{equation}
\Omega_{gl}=0 \quad \mbox{for CS},\quad \Omega_{gl}=-\frac{9}{64\pi^2}(gB)^2-2\frac{(gB)^2}{g^2} \quad \mbox{for CF}
\end{equation}
respectively.

Secondly, we wish to estimate the free energies, $\Omega_{q}$ of quarks. 
Before doing so,
we give a brief review of the color ferromagnetic phase
(see \cite{cf3} for more details). 
As has been shown, one-loop effective potential of
color magnetic field, $B$, shows not only a non-trivial minimum, but also
the presence of imaginary part for $gB\neq 0$,
$V(gB)=\frac{9}{32\pi^2}(gB)^2(\log(gB/\Lambda^2)-1/2)-\frac{i}{8\pi}(gB)^2$.
The real part of $V(gB)$ shows the spontaneous generation of 
the color magnetic field.
(We may assume that the direction of the field in the color space points
to $\lambda_3$.) 
But the presence of the imaginary part of $V(gB)$
indicates that the state is unstable. 
In other words the state is not true minimum in energy. Actually,
There exist unstable modes of gauge fields with
$\lambda_3$ color
charges.
These unstable modes occupy the lowest Landau level ($n=0$) 
with spins parallel to $B$ and have 
spectrum, $E_g^2=k^2+2gB(n+1/2)-2gB=k^2-gB<0$, where
$k$ represents a momentum parallel to the spatial direction of the
magnetic field. Among them,
the modes with $E_g(k=0)$ whose amplitudes increase most rapidly,
make a stable vacuum. 
As we have shown in previous papers, the unstable modes of the gauge
fields with $k=0$, which are spatially two-dimensional ones, 
condense to make stable
fractional quantum Hall states: Repulsive self-interaction of the gluons 
occupying the Lowest Landau
level leads to
the quantum Hall state even if the interaction is 
small. 
This formation of the quantum Hall
state\cite{qh} is similar to the case that two-dimensional electrons under
strong magnetic field form fractional quantum Hall states due to the Coulomb repulsion.
Since the condensate has the color charges,
the formation of the quantum Hall states is possible 
only when the quark matter is present;
the quarks supply the color charges for the stable vacuum.
Therefore, the color ferromagnetic phase 
is a possible state of the quark matter with sufficiently large baryon number
density so as for the loop approximation to be valid.
The gluon condensations leading to the quantum Hall states
produce a sufficiently large magnetic mass to the unstable modes
to stabilize the color ferromagnetic state. 
In this way, it turns out that the state in true minimum of the effective potential 
is a stable CF state which gains the condensation energy of the gluons, $-2B^2$.
 
Here, we note
that there is one unknown dimensional parameter in QCD 
associated with a renormalization scale, $\Lambda$ in $V(gB)$. 
In our case $gB$ is such a parameter.
Although the reliable value of $gB$ 
has not yet been determined phenomenologically, 
we expect that it has a typical scale of QCD. 
Here, we assume that the value of
$gB$ is around $(200\mbox{MeV})^2$.

 Now, we turn to calculate the free energies of quarks in the CF phase.
Quarks in SU(3) gauge theory are in a color triplet; 
$q=(q_{\rm r},q_{\rm g},q_{\rm b})$.
Among them, quarks, $q_{\rm r}$ and $q_{\rm g}$, occupy Landau levels under the
color magnetic field, $B\propto \lambda_3$.
Spectra of such quarks are
given by $E_q^2=m_f^2+k^2+gB(n+1/2+s_z)$ where $s_z=\pm 1/2$ represents a
spin contribution and $m_f$ is the quark mass with flavor, $f$. We take $m_{\rm u,d}=0$ and 
$m_{\rm s}=100\mbox{MeV}\sim 300\mbox{MeV}$. 
Each Landau level has 
the degeneracy per unit area being given by $gB/4\pi$.
The spectra of the quarks, $q_{\rm r}$ and $q_{\rm g}$, are identical
to each other.
%
Hence, the number density, $n_f$, and the energy density, $\rho_f$, of each
flavor of quarks are given by
\begin{align}
 n_f =& 
 =\frac{gB}{4\pi^2} \sqrt{\mu_f^2-m_f^2}+\cdots \ ,
\hspace{2em} \rho_f  =  
\frac{gB}{8\pi^2}\left(\mu_f\sqrt{\mu_f^2-m_f^2}
+m_f^2\ln\frac{\mu_f+\sqrt{\mu_f^2-m_f^2}}{m_f}\right)+\cdots \ , 
\end{align}
where we have explicitly shown only the contributions of the lowest Landau level;
we should understand that contributions of higher Landau levels are present in the above formulae.
$\mu_f$ denotes the chemical potential of
quarks, with the color types, $q_{\rm r}$ and $q_{\rm g}$. 
We should note that the chemical potential of $q_{\rm r}$ is the same as the 
one of $q_{\rm g}$, 
since we have an exchange symmetry between $q_{\rm r}$ and $q_{\rm g}$.
On the other hand, spectra of the quark, $q_{\rm b}$, are
given by $\sqrt{m_f^2+\vec{k}^2}$. Namely, the quarks do not
couple with the magnetic field.
Then, the number density, $\tilde{n}_f$, and the energy density, 
$\tilde{\rho}_f$, of the quark with the color type, $q_{\rm b}$, are given, respectively, by
\begin{align}
 \tilde{n}_{f} =& \frac{(\tilde{\mu}_f^2-m_f^2)^{3/2}}{3\pi^2}\ , 
\hspace{2em} \tilde{\rho}_{f} = \frac{2\tilde{\mu}_f^2-m_f^2}{8\pi^2}
\tilde{\mu}_f\sqrt{\tilde{\mu}_f^2-m_f^2} -\frac{m_f^4}{8\pi^2}
\ln\frac{\tilde{\mu}_f+\sqrt{\tilde{\mu}_f^2-m_f^2}}{m_f} \ , 
\end{align}
where $\tilde{\mu}_f$ is the chemical potential of the quarks, $q_{\rm b}$.

In the following, we should take into account three important conditions\cite{alford} 
which must be satisfied by the quarks and electrons 
in a realistic situation, namely, in neutron stars.
These are the conditions of color neutrality, electric neutrality and
beta-equilibrium,
\begin{align}
\label{color}
&n_{\rm u}+n_{\rm d}+n_{\rm s}=\tilde{n}_{\rm u}+\tilde{n}_{\rm d}+
\tilde{n}_{\rm s} & \mbox{(color neutrality)}, \\
\label{electric}
& \frac{2}{3}(2n_{\rm u}+\tilde{n}_{\rm u})-\frac{1}{3}(2n_{\rm d}+2n_{\rm s}+\tilde{n}_{\rm d}+\tilde{n}_{\rm s})-n_e=0  &
\mbox{(electric neutrality)}, \\
\label{beta}
&\mu_{\rm s}=\mu_{\rm d}=\mu_{\rm u}+\mu_e  \quad \mbox{and} \quad \tilde{\mu}_{\rm s}=\tilde{\mu}_{\rm d}=\tilde{\mu}_{\rm u}+\mu_e &
 \mbox{(beta-equilibrium)},
\end{align}
where $n_e=\mu_e^3/3\pi^2$ and $\mu_e$ denote the number density of 
electrons and their chemical potential, respectively. 
The color neutrality condition comes
from the average of each color charge vanishing, 
$\langle\lambda_3\rangle=\langle\lambda_8\rangle=0$.
The electric neutrality condition comes from the fact that 
the total electric charge density in neutron star should vanish. 
(Here we do not consider a possibility of mixing phase.)
Note that the number density of each flavor 
$\bar{n}_f$ is given by, 
$\bar{n}_f=2n_f+\tilde{n}_f$.
Furthermore, the beta-equilibrium
condition comes from u, d and s quarks transforming into each other
in neutron stars such as $d \,\,\,\mbox{or s} \to u+e$, or its
inverse process due to the weak interactions. These conditions reduce 
many independent variables to only one such as the baryon chemical potential.

Using these number densities and energy densities,
we can calculate the free energy of the quark matter 
in the color ferromagnetic phase,
\begin{eqnarray}
\Omega_q^{\rm CF}&=&2\rho_{\rm u}+\tilde{\rho_{\rm u}}+2\rho_{\rm d}
+\tilde{\rho_{\rm d}}+2\rho_{\rm s}+\tilde{\rho_{\rm s}}+\rho_e
-\sum_{f=\rm u,d,s}(2\mu_f n_f+\tilde{\mu}_f \tilde{n}_f)
-\mu_e n_e \nonumber  \\
&=&2\rho_{\rm u}+\tilde{\rho}_{\rm u}+2\rho_{\rm d}+\tilde{\rho}_{\rm d}
+2\rho_{\rm s}+\tilde{\rho}_{\rm s}+\rho_e-\mu_B
n_B
\end{eqnarray}
\noindent where
the second line of the equation has been derived using the conditions,
eqs.(\ref{color})-(\ref{beta}).
 $\mu_B$ and $n_B$ are the baryon chemical potential and the baryon number
density. ($\rho_e=\mu_e^4/4\pi^2$ is the energy
density of electrons.)
These can be represented such that,
\begin{align}
\mu_B &=\frac{2(\mu_{\rm u}+\mu_{\rm d}+\mu_{\rm s})
+\tilde{\mu}_{\rm u}+\tilde{\mu}_{\rm d}+\tilde{\mu}_{\rm s}}{3}, 
\\ 
 n_B&=\frac{\bar{n}_{\rm u}+\bar{n}_{\rm d}+\bar{n}_{\rm s}}{3}.
\end{align}
As we have mentioned, this free energy as well as the other
physical quantities can be expressed
simply in terms of the baryon chemical potential, $\mu_B$, 
using the above three conditions.  
 
In this way, we can calculate the free energy of the quarks in the
color ferromagnetic phase.
When the baryon number density becomes larger, 
quarks occupy higher Landau levels.
We can find numerically that
as quarks start to occupy higher Landau levels than the lowest Landau level, 
the free energy rapidly approaches to that
of the normal quark gas with no magnetic field;
$\Omega_q^{\rm CF}/\Omega_q^{\rm NG}
\to 1 $ as $\mu \to \infty$,
as we have shown in the previous paper\cite{cf}.
This is because the ratio depends only on the
dimensionless quantity $gB/\mu^2$ in the limit.

In order to compare this free energy with those of the color
superconducting states (CFL and 2SC) and the normal quark gas
state (NG), we write down the free energies of these states\cite{alford},
\begin{align}
\Omega_q^{\rm CFL}&=\frac{-\mu_B^4+9\mu_B^2m_{\rm s}^2}{108\pi^2} {
 -\frac{m_{\rm s}^4}{32\pi^2}\left(1-12\ln\frac{3m_{\rm s}}{2\mu_B}\right)}
-\frac{\mu_B^2\Delta_{\rm CFL}^2}{3\pi^2} &  \\
\Omega_q^{\rm 2SC}&=\frac{-\mu_B^4+9\mu_B^2m_{\rm s}^2}{108\pi^2} {
-\frac{m_{\rm s}^4}{32\pi^2}\left(5-12\ln\frac{3m_{\rm s}}{2\mu_B}\right) }
-\frac{\mu_B^2\Delta_{\rm 2SC}^2}{9\pi^2} &  \\
\Omega_q^{\rm NG}&=\frac{-\mu_B^4+9\mu_B^2m_{\rm s}^2}{108\pi^2} {
-\frac{m_{\rm s}^4}{32\pi^2}\left(7-12\ln\frac{3m_{\rm s}}{2\mu_B}\right) } 
& 
\end{align}
where $\Delta_{\rm CFL}$ and $\Delta_{\rm 2SC}$ are 
the gap energies in the color superconducting phases, CFL and 2SC, respectively. 
The formulae have been obtained up to second order in
$m_{\rm s}^2/\mu_B^2$, with the assumption that $m_{\rm s}^2$ is of the same order 
as $\mu_B\Delta_c/3$ ($c={\rm CFL}$ or $c={\rm 2SC}$).  
Then, the last two terms in $\Omega^{\rm CFL}$ and $\Omega^{\rm 2SC}$
become of the same order of the magnitude. It turns out that
the di-quark pairing effects in the superconducting states reduce 
the energies by $(\mu_B\Delta_c/3)^2\sim m_{\rm s}^4$ from the energy of NG. 

We should mention that the free energies obtained above
are reliable only for the quark matter with the small gauge coupling constant, in other words,
with sufficiently large baryon number density.
Quarks interact weakly with each others.
Hereafter we apply the free energies for quark matters with
realistic baryon number density such as several times of normal nuclear density.
It is expected that such quark matters are present in neutron stars.

Theoretically, the values of the gap parameters have much ambiguity,
so that we consider a range of the parameter
such as $10\,\mbox{MeV}<\Delta_c<100\,\mbox{MeV}$, supposing
$\Delta_{\rm CFL}\simeq \Delta_{\rm 2SC}$. We also consider a range of the strange
quark mass such as $100\,\mbox{MeV}<m_{\rm s}<300\, \mbox{MeV}$. 

In Fig.1 we show the free energy densities of the phases
in terms of the baryon
chemical potential, $\mu_B$ relative to the free energy density of NG phase,
i.e.~$\Delta \Omega=\Omega_q^c-\Omega_q^{\rm NG}$.
We take $gB=(200\mbox{MeV})^2$ and a large strange quark
mass, $m_{\rm s}=275$MeV along with a large gap parameter $\Delta=80$MeV
to reduce the free energies of CFL or 2SC phases.
(We have taken a value of the gauge coupling constant, $g^2/4\pi=0.3$.)
Obviously, the color ferromagnetic phase
is the most favored in the region depicted in Fig.1.
With the increase of the baryon chemical potential, higher Landau
levels are occupied and the CF free energy approaches to 
that of the normal quark matter; $\Omega_q^{\rm CF}\simeq \Omega_q^{\rm NG}-b(gB)^2$.
Eventually, CFL phase 
becomes energetically more favored than the ferromagnetic phase
in the extremely large chemical potential as can be seen in Fig.1.

Physically, the energy density of the quarks 
($\Omega_q^{\rm CF}\sim -O(gB\mu^2)$) in the magnetic field 
is much smaller than the
 energy density  of the normal quark gas 
($\Omega_q^{\rm NG}\sim -O(\mu^4)$), since all of quarks 
occupy the lowest Landau level in the 
limit of large magnetic field.
Even if the pairing effects 
in the color superconducting states are taken into account,
this situation is not altered. 
The pairing 
effects reduce the energy density of the normal quark gas 
only by the order of $\Delta^2\mu^2$; 
$\Omega_q^{\rm CFL}\simeq \Omega_q^{\rm NG}-a\Delta^2\mu^2$ with a
positive numerical factor $a$.  
With increasing the quark chemical
potential, $\mu$, 
quarks become occupying higher Landau levels
and the energy density approaches rapidly that of the normal quark
\vspace*{1em}

\noindent
\begin{minipage}{.45\textwidth}
\begin{center}
\setcaption{figure}
  \includegraphics[width=8.4cm,clip=yes]{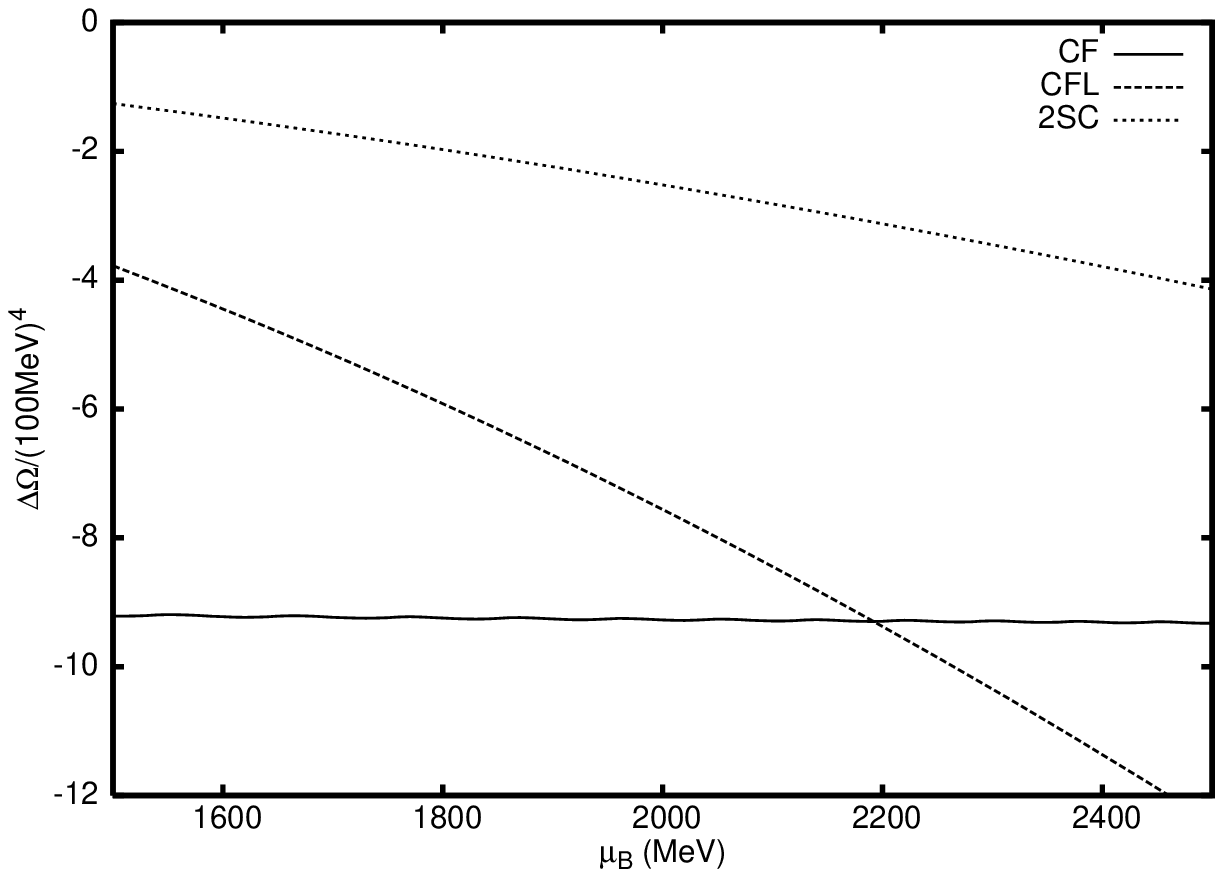}
  \caption{The free energy of various phases as a function of 
the chemical potential $\mu_B$, relative to the normal quark gas state.}
  \label{fig:freeenergy}
\end{center}
\end{minipage}
\hspace{2em}
\begin{minipage}{.5\textwidth}
\begin{center}
\setcaption{figure}
  \includegraphics[width=8.5cm,clip=yes]{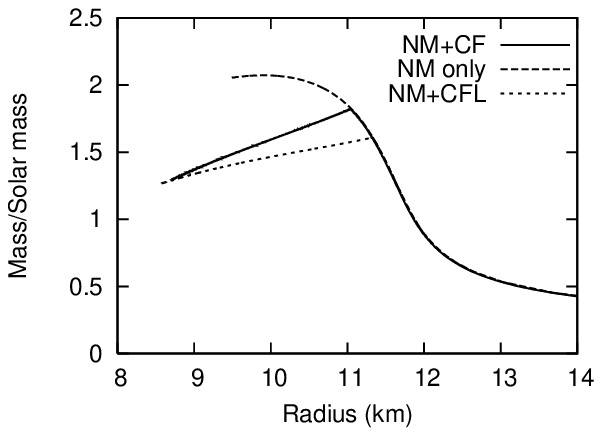}
  \caption{Mass-radius relation of the compact stars
in unit of the solar mass.}
  \label{fig:MR}
\end{center}
\end{minipage}
gas\cite{cf};
$\Omega_q^{\rm CF}\simeq \Omega_q^{\rm NG}-b(gB)^2$
with a positive numerical factor, $b\simeq 0.055 $.
This is because the limit of $\mu\to \infty$, is equivalent
to the limit of $gB\to 0$. Hence, 
in an intermediate value of $\mu$, $\Omega_q^{\rm CF}$ becomes larger
than $\Omega_q^{\rm CFL}$. This general argument implies that
as far as $b(gB)^2>a\Delta^2\mu^2$,
the color ferromagnetic phase is more favored, 
i.e.~$\Omega^{\rm CF}<\Omega^{\rm CFL}$ for small chemical potential
or large magnetic field.

In this way, the color ferromagnetic phase is more favored
than the CFL phase or 2SC phase, if the chemical potential, $\mu$,
satisfies, $b(gB)^2>a\Delta^2\mu^2$, but is sufficiently
large for the loop approximation to be valid.
Then, an intriguing question we should ask is \lq what 
is the critical baryon chemical potential
between hadron phase and the quark phase ?'.
In order to answer the 
question, we need to know precisely the free energy of gluons $\Omega_{gl}$, in the color
ferromagnetic phase. The critical point 
depends very sensitively on the value, $\Omega_{gl}$,
which is of the order of $(100\mbox{MeV})^4$.
Due to a forth power of an energy scale,
its contribution to the free energy is 
quite sensitive to the energy scale so that
it is difficult to obtain precisely the critical point.

Finally,
in Fig.2 we show the mass-radius relation of a 
hybrid neutron star composed of
a nuclear matter
and the quark matter in the CF phase
in its core.
The relation has been obtained by solving 
the Tolman-Oppenheimer-Volkoff equation
with the use of the equation of states of the quark matter
as well as the nuclear matter. 
As a nuclear matter, we have used an equation of state by
Akmal, Pandharipande, and Ravenhall (APR)\cite{apr}.
We have used tentatively $\Omega_{gl}=(180\mbox{MeV})^4$,
$gB=(200\mbox{MeV})^2$, $\Delta_c=80$MeV and $m_{\rm s}=275$MeV.
In Fig.2,
the dotted line represents the relation for a neutron star containing only
nuclear matter (NM).  The solid line for the hybrid star involving the 
CF quark matter inside of the nuclear matter. 
For comparison, we also show the relation (dashed line) of
a hybrid star composed of the CFL quark
matter with the same $\Omega_{gl}$ inside of the nuclear matter.
It turns out that the critical mass of the hybrid neutron
stars discussed in the paper is consistent with observations\cite{ob}. 
We will report the details in the near future.

\vspace*{2em}

A.~I.~and M.~O.~express thanks to the member of theory 
group in KEK for their hospitality.
This work was partially supported by Grants-in-Aid of the Japanese Ministry
of Education, Science, Sports, Culture and Technology (No. 13135218).



\begin{thebibliography}{99}
\bibitem{neutron}S.L. Shapiro and S.A. Teukolsky, Black Holes, White Dwarfs 
and Neutron Stars, John Wiley and Sons, (1983).
\bibitem{ob}J.M. Lattimer and M. Prakash, astro-ph/0405262.
\bibitem{color}K. Rajagopal and F. Wilczek, hep-ph/0011333.
\bibitem{cf}A. Iwazaki and O. Morimatsu, Phys. Lett. B 571 (2003) 61; \\
A. Iwazaki, O, Morimatsu, T. Nishikawa and M. Ohtani,
Phys. Lett. B 579 (2004) 347.
\bibitem{cf3}A. Iwazaki, O, Morimatsu, T. Nishikawa and M. Ohtani,
Phys. Rev. D 71 (2005) 034014.
\bibitem{savvidy}G.K. Savvidy, Phys. Lett. B 71 (1977) 133.\\
H. Pagels, Lecture at Coral Gables, Florida, 1978.
\bibitem{huang}M. Huang and I.A. Shovkovy, Phys. Rev. D 70 (2004) 094030.
\bibitem{nielsen}N.K. Nielsen and P. Olesen, Nucl. Phys. B 144 (1978) 376;
Phys. Lett. B 79 (1978) 304.
\bibitem{bor}A.O. Starinets, A.S. Vshivtsev and V.Ch. Zhukovskii,
  Phys. Lett. B 322 (1994) 403; \\
V. Skalozub and M. Bordag, Nucl. Phys. B 576 (2000) 430.
\bibitem{qh}S. Das Sarma and A. Pinczuk (Eds.),
Perspectives in Quantum Hall Effects, (Wiley, New York, 1997). 
\bibitem{alford}M. Alford and K. Rajagopal, JHEP 0206 (2002) 031; \\
M. Alford, M. Braby, M. Paris and S. Reddy, nucl-th/0411016.
\bibitem{apr}A. Akmal, V.R. Pandharipande and D.G. Ravenhall,
  Phys. Rev. C 58 (1998) 1804.
\end{thebibliography}
\end{document}